\documentclass[conference]{IEEEtran}
\IEEEoverridecommandlockouts

\usepackage{amsmath,amsfonts,amssymb}
\usepackage{tikz}
\usetikzlibrary{arrows.meta, positioning, calc}
\usepackage[siunitx,american]{circuitikz}
\usepackage[ruled,vlined]{algorithm2e}
\usepackage{algpseudocode}
\usepackage{cite}
\usepackage{amsmath,amssymb,amsfonts,bm}
\usepackage{graphicx}
\usepackage{balance}
\usepackage{textcomp}
\usepackage{xcolor}
\usepackage{multirow}
\def\BibTeX{{\rm B\kern-.05em{\sc i\kern-.025em b}\kern-.08em
    T\kern-.1667em\lower.7ex\hbox{E}\kern-.125emX}}

\newcommand\freefootnote[1]{%
  \let\thefootnote\relax%
  \footnotetext{#1}%
  \let\thefootnote\svthefootnote%
}

\renewcommand{\baselinestretch}{0.9} 
\begin{document}

\newcommand\mycommfont[1]{\small\ttfamily{#1}}
\SetCommentSty{mycommfont}

\newcommand\scalemath[2]{\scalebox{#1}{\mbox{\ensuremath{\displaystyle #2}}}}

\title{Design-Oriented Modeling of TSV Substrate Noise Coupling to Ring VCOs
}

\author{
    \IEEEauthorblockN{Ilias Exouzidis\textsuperscript{1,2}, Alberto Garcia-Ortiz\textsuperscript{3}, George Floros\textsuperscript{2,4}, Georgios Panagopoulos\textsuperscript{1}}
    \IEEEauthorblockA{\textsuperscript{1}Department of Electrical and Computer Engineering, National Technical University of Athens, Athens, Greece}
    \IEEEauthorblockA{\textsuperscript{2}Department of Electrical and Computer Engineering, University of Thessaly, Volos, Greece}
    \IEEEauthorblockA{\textsuperscript{3}Institute of Electrodynamics and Microelectronics, University of Bremen, Germany}
    \IEEEauthorblockA{\textsuperscript{4}Department of Electronic and Electrical Engineering, Trinity College Dublin, Dublin, Ireland}
    \IEEEauthorblockA{elias\_exouzidis@mail.ntua.gr, agarcia@item.uni-bremen.de, florosg@tcd.ie, gepanago@mail.ntua.gr}
}

\maketitle

\begin{abstract}
Through-silicon vias (TSVs) enable dense vertical interconnects in 3D-IC and chiplet systems, but their metal–oxide–silicon structure introduces significant parasitic coupling paths that can degrade the spectral purity of sensitive RF blocks. This paper presents a compact, design-oriented methodology for assessing TSV-induced substrate noise in mixed-signal circuits. We derive a closed-form analytical three-port RLGC macromodel for a Signal–Ground TSV pair that explicitly exposes the substrate node. The methodology is validated using a three-stage Ring VCO designed in a 22 nm FD-SOI technology, where specific RF devices from the process design kit (PDK) provide direct access to the transistor substrate terminals for controlled noise injection. Multi-tone Harmonic Balance simulations in Spectre RF quantify the impact of TSV aggressors on the oscillator's output spectrum. The results indicate that an aggressor of 1 GHz, 0.5 V$_{pp}$ induces a primary sideband spur of -35.2 dBc. Sensitivity characterization reveals that the magnitude of these sideband spurs increases monotonically with the aggressor amplitude. Furthermore, frequency sweeps demonstrate a low-pass coupling response, where the induced spur magnitude decreases from -20.2 dBc at 500 MHz to -33.1 dBc at 2 GHz. 
\end{abstract}


\section{Introduction} 

Three-dimensional (3D) integration utilizing through-silicon vias (TSVs) has emerged as a critical technology for high-performance heterogeneous systems. These vertical interconnects offer reduced interconnect latency, increased bandwidth density, and improved form-factor scalability. However, the physical structure of the TSV, which typically consists of a copper-filled cylinder isolated from the silicon substrate by a thin dielectric liner, inherently forms a metal-oxide-semiconductor (MOS) capacitor \cite{katti2010_electrical_tsv}. This parasitic capacitance facilitates significant electrical coupling between the vertical interconnects and the conductive substrate, creating pathways for digital switching noise to inject into the common substrate and propagate to sensitive analog and RF blocks \cite{7849155}. \freefootnote{*This work was funded under the COIN-3D project, which has received funding from the European Union’s Horizon Europe research and innovation program under grant agreement No. 101159667.}

To address these signal integrity challenges, extensive research has focused on the electrical modeling of TSVs. This includes analytical RLGC formulations that account for skin effect, eddy currents, and depletion capacitance as functions of TSV geometry and doping profiles \cite{katti2010_electrical_tsv,kim2011hfscalabletsv}. Recent advances have extended these models to consider Signal--Ground (SG) pairs, providing closed-form expressions for insertion loss and crosstalk that align well with electromagnetic (EM) field solvers while drastically reducing simulation runtime \cite{gharib2024analytical}. In parallel, significant effort has been devoted to analyzing mitigation strategies, such as guard rings and shielding TSVs, to isolate victim circuits \cite{cho2011tsvguardring}. Notably, Lim et al. expanded on this by assessing TSV-to-oscillator noise coupling and evaluating specific mitigation techniques, combining EM-based extraction with phase-noise evaluation for LC-VCOs \cite{lim2018tsvnoise}.

Despite these modeling advances, a gap remains in bridging interconnect-level parasitics with transistor-level circuit performance. Although recent studies have successfully employed 20~GHz LC-VCOs as monitors to validate substrate coupling flows in planar RF-CMOS SoCs \cite{karipidis2023simulation}, the specific impact of vertical TSV-induced noise on digitally-intensive ring voltage-controlled oscillators (Ring VCOs) in Fully Depleted Silicon-on-Insulator (FD-SOI) technologies remains under-explored. 

To address this challenge, this paper proposes a compact, design-oriented methodology to quantify TSV-induced substrate noise in Ring VCOs. The primary contributions of this work are summarized as follows. \textit{First}, we propose an analytical three-port RLGC model for a SG TSV pair that explicitly exposes a substrate port, enabling direct modeling of substrate coupling mechanisms within circuit-level simulations. \textit{Second}, the proposed methodology is validated at the transistor level using a three-stage, current-starved Ring VCO designed in a 22~nm FD-SOI process. By utilizing specific RF devices from the PDK to access the explicit substrate terminals, we establish a direct co-simulation environment between the TSV model and the oscillator core. \textit{Finally}, we quantify TSV-induced spectral degradation in the VCO, characterize the sensitivity of sideband spurs to aggressor amplitude and analyze the frequency-dependent filtering of the TSV-substrate network.

The remainder of the paper is organized as follows: Section~\ref{sec:tsv_model} details the analytical derivation of the three-port TSV model and its S-parameter extraction. Section~III describes the 22~nm FD-SOI Ring VCO design and the explicit substrate biasing scheme. Section~IV presents the simulation results with and without TSV-induced substrate excitation, analyzing spur magnitude, sensitivity to noise magnitude and frequency response. Finally, conclusions are drawn in Section~V.

\section{Analytical 3-Port TSV Signal-Ground Model}
\label{sec:tsv_model}

To capture TSV-induced substrate-borne disturbances for mixed-signal 3D IC assessment, we employ a closed-form analytical RLGC macromodel for a SG TSV pair. Crucially, unlike standard two-port models that lump substrate effects into shunt losses, this work explicitly exposes the common silicon node as a third terminal (Port 2). This allows for the direct injection of noise into the body potential of active devices during circuit co-simulation.
The modeling approach adapts the methodologies in \cite{gharib2024analytical,katti2010_electrical_tsv} to yield a frequency-dependent impedance matrix $\mathbf{Z}(s)$, which is subsequently converted to S-parameters.

\begin{figure}[!t]
    \centering
    \includegraphics[width=0.8\linewidth]{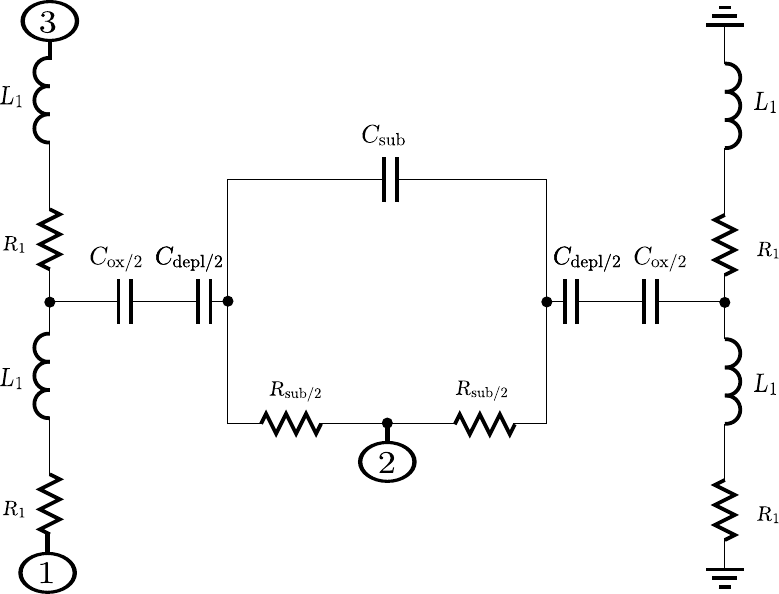}
    \caption{Lumped-element RLGC topology of the SG TSV pair with explicit substrate terminal (Node 2).}
    \vspace{-3 mm}
    \label{fig:tsv_rlgc}
\end{figure}

\textit{Series Resistance:} The TSV series resistance consists of a DC component due to the finite conductivity of copper and an AC component arising from high-frequency current crowding (skin effect). The DC resistance is modeled as:
\begin{equation}
  R_{\mathrm{DC}} = \frac{\rho_{\mathrm{Cu}} h_{\mathrm{tsv}}}{\pi r_{\mathrm{tsv}}^{2}},
\end{equation}
where $\rho_{\mathrm{Cu}}$ is the copper resistivity, $h_{\mathrm{tsv}}$ is the TSV height, and $r_{\mathrm{tsv}}$ is the TSV radius. The skin depth $\delta(f)$ characterizes the penetration of the RF current into the conductor, which leads to the frequency-dependent AC resistance approximation:
\begin{equation}
  \delta(f) = \sqrt{\frac{\rho_{\mathrm{Cu}}}{\pi f \mu_r \mu_0}},
\qquad
  R_{\mathrm{AC}}(f) = \frac{\rho_{\mathrm{Cu}} h_{\mathrm{tsv}}}{2 \pi r_{\mathrm{tsv}} \,\delta(f)} ,
\end{equation}
where $f$ is the excitation frequency, $\mu_0$ is the vacuum permeability, and $\mu_r$ is the relative permeability of copper.
To ensure a continuous model valid from DC to RF, the total resistance is approximated by the quadrature sum:
\begin{equation}
  R_{\mathrm{TSV}}(f) = \sqrt{R_{\mathrm{DC}}^{2} + R_{\mathrm{AC}}^{2}(f)}.
\end{equation}
In the lumped network shown in Fig.~\ref{fig:tsv_rlgc}, this resistance is divided equally between the two vertical segments, i.e., $R_{1}(f)=R_{\mathrm{TSV}}(f)/2$.

\textit{Capacitive and Conductive Substrate Paths:} The TSV liner and surrounding silicon form a MOS-like radial stack. The liner contributes an oxide capacitance between the Cu core and the silicon:
\begin{equation}
  C_{\mathrm{ox}} =
  \frac{2 \pi \varepsilon_{\mathrm{ox}} \varepsilon_{0} h_{\mathrm{tsv}}}
       {\ln\!\left( \frac{r_{\mathrm{tsv}} + t_{\mathrm{ox}}}{r_{\mathrm{tsv}}} \right)} ,
\end{equation}
where $\varepsilon_{\mathrm{ox}}$ is the relative permittivity of the oxide liner, $\varepsilon_0$ is the vacuum permittivity, and $t_{\mathrm{ox}}$ is the liner thickness. 
In the substrate, the radial electric field creates a depletion region of width $W_d$ around the via, yielding the corresponding depletion capacitance \cite{katti2010_electrical_tsv}:
\begin{equation}
  W_d =
  \sqrt{\frac{4\varepsilon_{\mathrm{si}}\varepsilon_{0}V_T
  \ln\!\left(\frac{N_A}{n_i}\right)}
  {qN_A}},
\end{equation}
where $\varepsilon_{\mathrm{si}}$ is the relative permittivity of silicon, $V_T=kT/q$ is the thermal voltage, $N_A$ is the acceptor doping concentration, and $n_i$ is the intrinsic carrier density.
\begin{equation}
  C_{\mathrm{d}} =
  \frac{2 \pi \varepsilon_{\mathrm{si}} \varepsilon_{0} h_{\mathrm{tsv}}}
       {\ln\!\left( \frac{r_{\mathrm{tsv}} + t_{\mathrm{ox}} + W_d}
                         {r_{\mathrm{tsv}} + t_{\mathrm{ox}}} \right)} ,
\end{equation}
Although $W_d$ varies non-linearly with the transmitted bit, this effect is negligible for RF signals and assumed constant \cite{bamberg2022_3d}.
Beyond the vertical stack, lateral coupling between the signal TSV and the adjacent ground TSV is captured by modeling the substrate as a lossy dielectric medium between parallel wires. This yields a shunt capacitance $C_{\mathrm{si}}$ and conductance $G_{\mathrm{si}}$ \cite{gharib2024analytical}:
\begin{equation}
  C_{\mathrm{si}} =
  \frac{\pi \varepsilon_{\mathrm{si}} \varepsilon_{0} h_{\mathrm{tsv}}}
       {\operatorname{acosh}\!\left( \frac{p_{\mathrm{tsv}}}{2 r_{\mathrm{tsv}}} \right)},
  \quad
  G_{\mathrm{si}} =
  \frac{\pi \sigma_{\mathrm{si}} h_{\mathrm{tsv}}}
       {\operatorname{acosh}\!\left( \frac{p_{\mathrm{tsv}}}{2 r_{\mathrm{tsv}}} \right)} ,
\end{equation}
where $p_{\mathrm{tsv}}$ is the TSV pitch and $\sigma_{\mathrm{si}}=1/\rho_{\mathrm{si}}$ is the silicon conductivity.

\textit{Series Inductance:} The TSV self-inductance arises from the magnetic energy stored around the current-carrying via. A Grover-type closed-form approximation for a cylindrical conductor is used \cite{katti2010_electrical_tsv}:
\begin{equation}
\begin{split}
  L_{\mathrm{TSV}} &=
  \frac{\mu_{0} \mu_{r} h_{\mathrm{tsv}}}{2 \pi}
  \Bigg[
    \ln\!\left(
      \frac{h_{\mathrm{tsv}}}{r_{\mathrm{tsv}}}
      + \sqrt{1 + \left(\frac{h_{\mathrm{tsv}}}{r_{\mathrm{tsv}}}\right)^{2}}
    \right)
\\[-2pt] &\qquad
    + \frac{r_{\mathrm{tsv}}}{h_{\mathrm{tsv}}}
    - \sqrt{1 + \left(\frac{r_{\mathrm{tsv}}}{h_{\mathrm{tsv}}}\right)^{2}}
  \Bigg] ,
\end{split}
\end{equation}
The inductance is distributed symmetrically in the lumped network using $L_{1}=L_{\mathrm{TSV}}/2$.

\begin{figure}[!t]
  \centering
  \includegraphics[width=\linewidth]{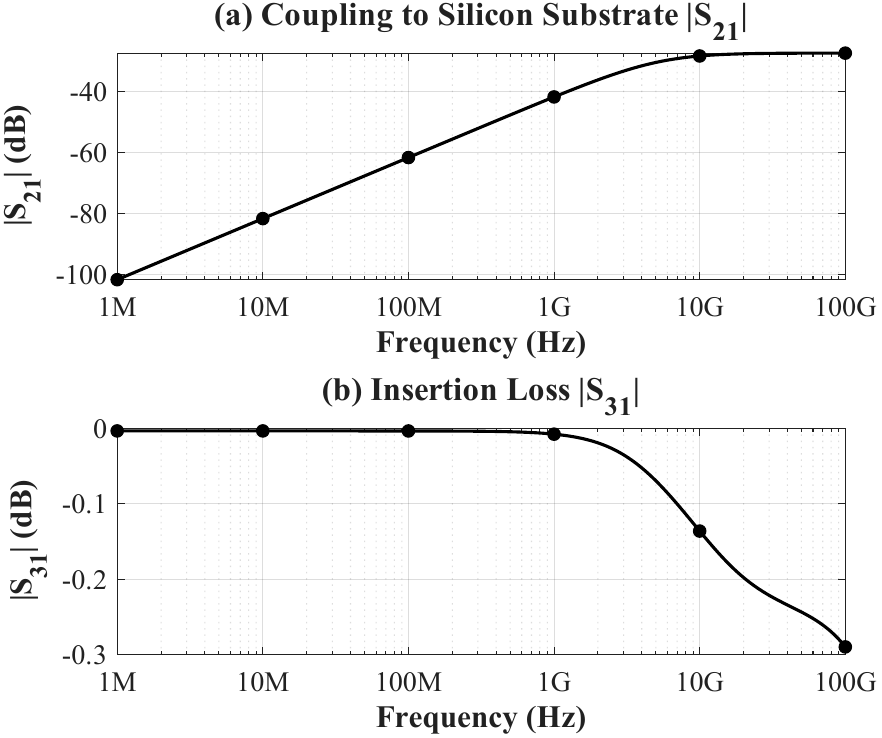}
  \caption{(a) Magnitude of the coupling coefficient to the substrate $|S_{21}|$. (b) Insertion loss of the TSV SG pair $|S_{31}|$.}
  \vspace{-3 mm}
  \label{fig:tsv_s_params}
\end{figure}

\textit{Three-port assembly and S-parameters:} The RLGC elements are assembled into a three-port lumped network presented in Fig.~\ref{fig:tsv_rlgc}. Port~1 corresponds to the bottom node of the signal TSV, Port~3 to the top node on the upper tier, and Port~2 to the effective substrate node.
The model is instantiated using typical geometrical parameters for 3D-IC processes \cite{bamberg2022_3d}: TSV height $h_{\mathrm{tsv}}=50\,\mu\mathrm{m}$, radius $r_{\mathrm{tsv}}=2.5\,\mu\mathrm{m}$, pitch $p_{\mathrm{tsv}}=40\,\mu\mathrm{m}$, and oxide liner thickness $t_{\mathrm{ox}}=0.5\,\mu\mathrm{m}$. The material properties correspond to standard process values, including oxide permittivity $\varepsilon_{\mathrm{ox}}=3.9$, copper resistivity $\rho_{\mathrm{Cu}}=1.68\times10^{-8}\,\Omega\cdot\mathrm{m}$, and a p-type silicon substrate with acceptor doping $N_A=1.2\times10^{15}\,\mathrm{cm}^{-3}$ (resulting in $\rho_{\mathrm{si}}\approx 0.12\,\Omega\cdot\mathrm{m}$). Standard physical constants ($\varepsilon_0, \mu_0, V_T$) are assumed throughout.
Closed-form expressions for the impedance parameters $Z_{ij}(s)$ are derived from this network and evaluated over a frequency grid from $1~\mathrm{MHz}$ to $100~\mathrm{GHz}$ to construct $\mathbf{Z}(s)$. The corresponding scattering parameters are computed as:
\begin{equation}
  \mathbf{S}(s) =
  \bigl( \mathbf{Z}(s) - Z_{0} \mathbf{I} \bigr)
  \bigl( \mathbf{Z}(s) + Z_{0} \mathbf{I} \bigr)^{-1},
  \label{eq:ZtoS}
\end{equation}
where $Z_{0}=50\,\Omega$ is the reference impedance and $\mathbf{I}$ is the identity matrix. The resulting $\mathbf{S}(s)$ is exported as a Touchstone file in order to be used for simulation in Cadence Virtuoso.
The magnitudes of $S_{21}$ and $S_{31}$ are shown in Figure~\ref{fig:tsv_s_params}. $|S_{31}|$ represents the signal insertion loss, while $|S_{21}|$ quantifies the coupling from the signal TSV into the silicon substrate. As seen in Figure~\ref{fig:tsv_s_params}a, the coupling to the substrate increases significantly with frequency, reaching approximately -30 dB at 10 GHz, which highlights the risk of high-frequency noise injection.

\section{Ring VCO Design}
\label{sec:ringvco}

To investigate the impact of TSV-induced substrate noise, we designed a three-stage Ring VCO in a 22\,nm FD-SOI process. The oscillator employs a current-starved topology, comprising three identical CMOS inverter stages. Frequency tuning is achieved by modulating the effective resistance of the inverter stages via the control voltage $V_{\mathrm{ctrl}}$, which adjusts the current supplied by the header and footer devices. The inverter stages are sized to target a center frequency of approximately 10.87~GHz, with device dimensions $(W/L)_n = 2.4\,\mu\mathrm{m}/40\,\mathrm{nm}$ and $(W/L)_p = 3.6\,\mu\mathrm{m}/50\,\mathrm{nm}$.

Linearization of the tuning characteristic is critical, as standard current-starved topologies often compress and saturate at low control voltages. To mitigate this, a feedback transistor ($M_5$) is introduced to regulate the operating point of the delay stages. This feedback mechanism prevents premature frequency saturation and linearizes the $f_{\mathrm{osc}}$ vs. $V_{\mathrm{ctrl}}$ curve, resulting in a stable VCO gain across the tuning range of 10.3--11.5~GHz.

Back-gate body-biasing is employed as an additional degree of freedom to shape the tuning behavior and stabilize the gain of the VCO. The PMOS devices are biased at $V_{\mathrm{bp}} = 0\,\mathrm{V}$, while the core NMOS devices are biased at $V_{\mathrm{bn}} = 0.8\,\mathrm{V}$. An exception is made for the NMOS control transistor $M_1$, located in the current-starving network, which utilizes a forward back-gate bias of $V_{\mathrm{bn}} = 2\,\mathrm{V}$. This higher bias lowers its effective threshold voltage, ensuring the device remains conductive even at low $V_{\mathrm{ctrl}}$ levels, thereby preventing premature saturation.

The primary objective of this design is to serve as a monitor for substrate coupling. Consequently, the inverter stages are implemented using specific RF devices from the GlobalFoundries 22FDX PDK RF library. Unlike standard digital cells, these devices provide explicit terminals for the local substrate (SUB) and Deep N-Well (DNW). This feature allows the explicit substrate port (Port 2) of our TSV macromodel to be connected directly to the local substrate of the VCO transistors, enabling the controlled injection and observation of substrate disturbances  in the simulation environment.

\begin{figure}[!b]
    \centering
    \vspace{-3 mm}
    \includegraphics[width=\linewidth]{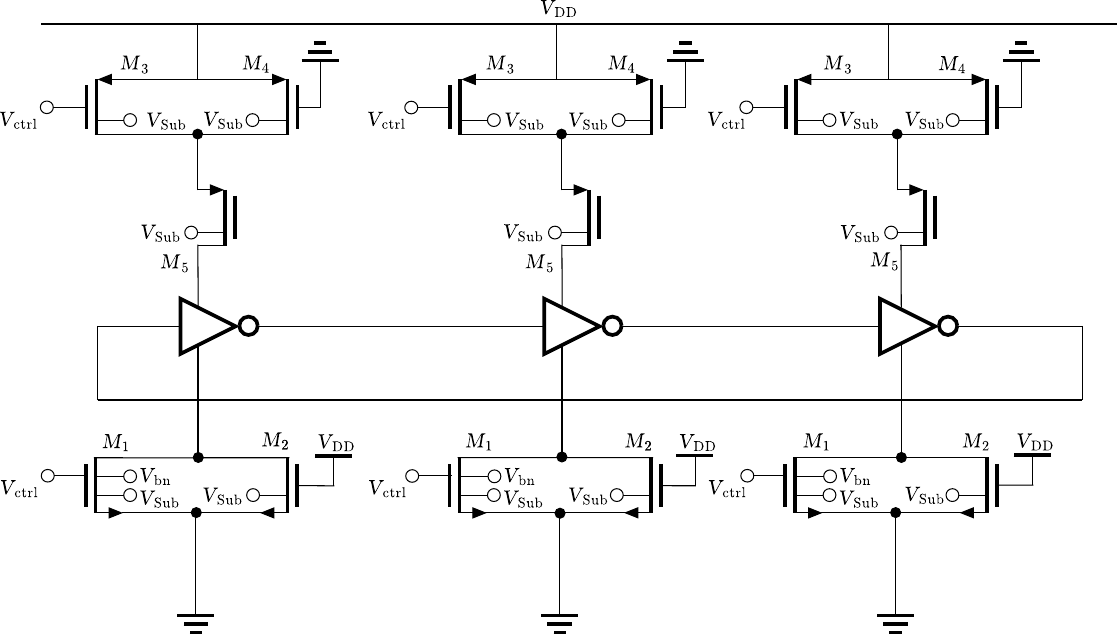}
    \caption{Schematic of the three-stage Ring VCO design.}
    \label{fig:ringvco_topology}
\end{figure}

\section{Results and Discussion}
\label{sec:results}

The impact of TSV-induced substrate noise on the VCO performance was quantified using multi-tone Harmonic Balance (HB) simulations in the Cadence Spectre RF environment. To ensure robust convergence of the autonomous circuit prior to the frequency-domain solution, a transient assist period (\texttt{tstab}) of 20 ns was employed. The HB analysis was configured with an order of 10 harmonics for both the fundamental ($f_{\text{osc}}$) and aggressor ($f_{\text{agg}}$) frequencies under the \texttt{conservative} solver accuracy mode in order to effectively capture mixing products and detect low-level spurs. The accuracy metric for the analysis is the relative magnitude in dBc, defined as the difference between the amplitude of the carrier signal and the substrate intermodulation spur.

\subsection{VCO Spectrum Analysis}
\begin{figure}[!b]
  \centering
  \vspace{-3mm}
  \includegraphics[width=\linewidth]{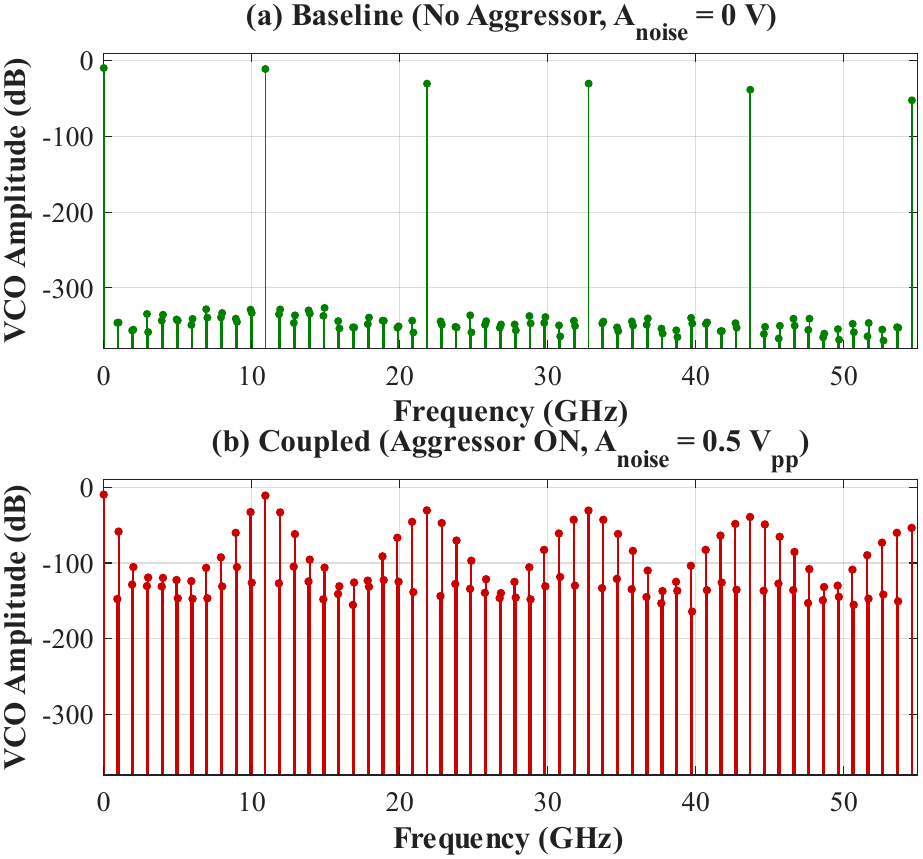} 
  \caption{(a) Baseline spectrum with inactive TSV aggressor ($A_{noise}=0$ V), showing the clean fundamental oscillation at $f_{osc} = 10.917$ GHz. (b) Coupled spectrum under the influence of a 1 GHz, 0.5 V$_{pp}$ TSV aggressor, revealing distinct sideband spurs due to Substrate Coupling.}
  \label{fig:vco_comparison}
\end{figure}

The 22 nm FD-SOI Ring VCO exhibits a free-running frequency $f_{\text{osc}}$ of 10.917 GHz at a control voltage $V_{\text{ctrl}}$ of 400 mV, with a nominal output power of -11.02 dB. To establish a reference for coupling analysis, the baseline spectrum was first simulated with the TSV aggressor inactive, meaning the noise amplitude $A_{\text{noise}}$ was set to 0 V. As shown in Fig.~\ref{fig:vco_comparison}a, the resulting spectrum is clean, and the intermodulation sideband spurs at $n f_{\text{osc}} \pm m f_{\text{agg}}$ are observed to be below -328 dB. This confirms that the simulation setup does not introduce intrinsic spectral artifacts.

Subsequently, a sinusoidal aggressor signal was injected into the Signal TSV port of the proposed macromodel. The explicit substrate port of the TSV model is connected directly to the common substrate node SUB of the Ring VCO. This node corresponds to the substrate terminal of the transistors provided by the PDK's RF library, thereby coupling the injected noise directly to the p-well of every active device in the design.

For the coupled spectrum shown in Fig.~\ref{fig:vco_comparison}b, the aggressor frequency was fixed at $f_{\text{agg}} = 1$ GHz with a peak-to-peak amplitude $A_{\text{noise}}$ of 500 mV. The injected noise propagates through the shared substrate network and mixes with the carrier, generating a series of distinct sideband spurs with significant amplitude. While the spectral noise floor rises across the band, the dominant interference manifests as a discrete first-order sideband spur at $f_{\text{osc}} + f_{\text{agg}}$, corresponding to a frequency of 11.917 GHz. This spur exhibits a level of -35.2 dBc, indicating that standard logic switching levels can significantly degrade the spectral purity of the oscillator.

\subsection{Sensitivity to Aggressor Amplitude}
To characterize the sensitivity of the VCO to noise magnitude, a swept Harmonic Balance simulation was performed. The aggressor amplitude $A_{\text{noise}}$ was swept from 100 mV to 700 mV while maintaining a constant frequency of 1 GHz. Fig.~\ref{fig:vco_sensitivity}a illustrates the magnitude of the first sideband spur, which is located at the frequency $f_{\text{osc}} + f_{\text{agg}}$, as a function of the aggressor amplitude.

As expected, the substrate coupling strength increases monotonically, with the induced spur magnitude rising from -36.1 dBc at 100 mV to -19.1 dBc at 700 mV. The response exhibits a consistent slope of approximately 6 dB per octave. This indicates that for standard digital logic levels, the coupling path operates in a linear regime, where the injected noise power scales proportionally with the aggressor's strength.

\subsection{Frequency-Dependent Coupling Response}
\begin{figure}[!t]
  \centering
  \includegraphics[width=\linewidth]{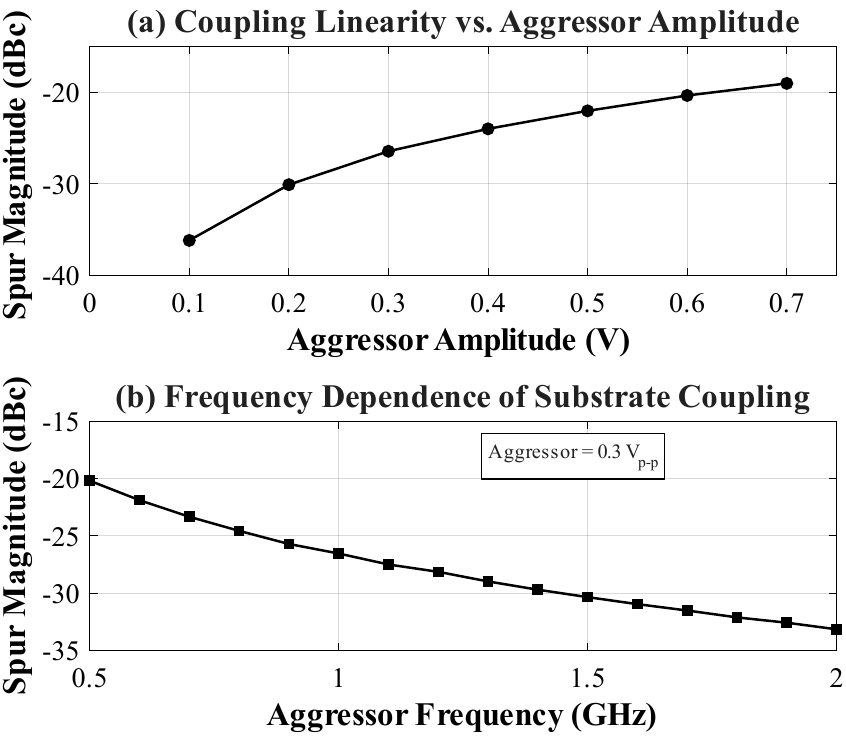}
  \caption{(a) Sensitivity analysis showing the magnitude of the first sideband spur ($f_{osc} + f_{agg}$) versus aggressor amplitude. (b) Frequency sweep showing the spur magnitude versus aggressor frequency ($A_{noise}=0.3$ V$_{pp}$).}
   \vspace{-3 mm}
  \label{fig:vco_sensitivity}
\end{figure}

Finally, the frequency dependence of the coupling path was evaluated using a swept Harmonic Balance analysis. The aggressor frequency $f_{\text{agg}}$ was swept from 0.5 GHz to 2 GHz, while the aggressor amplitude was maintained at a constant level of 300 mV$_{pp}$. Fig.~\ref{fig:vco_sensitivity}b illustrates the magnitude of the first sideband spur, located at $f_{\text{osc}} + f_{\text{agg}}$, as a function of the aggressor frequency.

The observed spur magnitude exhibits a significant decrease across the swept range. At 500 MHz, the spur level is $-20.2$ dBc. As the frequency increases to 2 GHz, the coupling weakens, resulting in a spur magnitude of -33.1 dBc. This trend reveals that high-frequency intermodulation products are subject to significantly greater attenuation than low-frequency ones. This roll-off is attributed to the inherent phase modulation properties of the oscillator. Since the magnitude of FM sidebands is inversely proportional to the modulation frequency ($1/f_{\text{agg}}$), the VCO becomes naturally less sensitive to phase disturbances as the aggressor frequency increases.

\section{Conclusion}
\label{sec:conclusion}

This work presented a design-oriented methodology to quantify TSV-induced substrate noise in Ring VCOs using a closed-form three-port TSV macromodel with an explicit substrate node. Validated on a 22~nm FD-SOI three-stage Ring VCO, the co-simulation framework confirmed that digital switching noise significantly degrades spectral purity. Specifically, a 1~GHz, 0.5~V$_{pp}$ aggressor induced a sideband spur of -35.2~dBc. Sensitivity analysis demonstrated a monotonic rise in coupling strength, reaching -19.1~dBc at a 700~mV aggressor amplitude, while frequency sweeps revealed a distinct low-pass behavior where the spur attenuated to -33.1~dBc at 2~GHz.


\bibliographystyle{IEEEtran}
\bibliography{bibliography}

@inproceedings{gharib2024analytical,
  title        = {An Analytical Model for High-Frequency Through Silicon Vias},
  author       = {Gharib, Mohamed Adel and Abdelzaher, Salma and Partin-Vaisband, Inna},
  booktitle    = {Proceedings of the Great Lakes Symposium on VLSI 2024 (GLSVLSI '24)},
  pages        = {282--286},
  year         = {2024}
}

@article{katti2010_electrical_tsv,
  author  = {Katti, Guruprasad and others},
  title   = {Electrical Modeling and Characterization of Through Silicon via for Three-Dimensional {IC}s},
  journal = {IEEE Transactions on Electron Devices},
  volume  = {57},
  number  = {1},
  pages   = {256--262},
  year    = {2010},
  month   = jan,
  doi     = {10.1109/TED.2009.2034508}
}

@article{lim2018tsvnoise,
  title        = {Modeling and Analysis of {TSV} Noise Coupling Effects on {RF} {LC}-{VCO} and Shielding Structures in {3D} {IC}},
  author       = {Lim, Jaemin and others},
  journal      = {IEEE Transactions on Electromagnetic Compatibility},
  journalabbr  = {IEEE Trans. Electromagn. Compat.},
  volume       = {60},
  number       = {6},
  pages        = {1939--1947},
  year         = {2018},
  month        = dec,
}

@article{cho2011tsvguardring,
  title        = {Modeling and Analysis of Through-Silicon Via ({TSV}) Noise Coupling and Suppression Using a Guard Ring},
  author       = {Cho, Jonghyun and others},
  journal      = {IEEE Transactions on Components, Packaging and Manufacturing Technology},
  journalabbr  = {IEEE Trans. Compon., Packag., Manufact. Technol.},
  volume       = {1},
  number       = {2},
  pages        = {220--233},
  year         = {2011},
  month        = feb,
}

@article{kim2011hfscalabletsv,
  title        = {High-Frequency Scalable Electrical Model and Analysis of a Through Silicon Via ({TSV})},
  author       = {Kim, Joohee and others},
  journal      = {IEEE Transactions on Components, Packaging and Manufacturing Technology},
  journalabbr  = {IEEE Trans. Compon., Packag., Manuf. Technol.},
  volume       = {1},
  number       = {2},
  pages        = {181--195},
  year         = {2011},
  month        = feb,
  doi          = {10.1109/TCPMT.2010.2101890},
}

@ARTICLE{7849155,
  author={Lu, Tiantao and others},
  journal={IEEE Transactions on Computer-Aided Design of Integrated Circuits and Systems}, 
  title={{TSV}-Based {3-D} {IC}s: Design Methods and Tools}, 
  year={2017},
  volume={36},
  number={10},
  pages={1593-1619},
  doi={10.1109/TCAD.2017.2666604}}

@article{karipidis2023simulation,
  title={Simulation of substrate coupling for mobile communications {SoC}--{A} 20 {GHz} {VCO} case study},
  author={Karipidis, S and others},
  journal={International Journal of Electronics and Communications (AE{\"U})},
  volume={161},
  pages={154548},
  year={2023},
  publisher={Elsevier}
}

@book{bamberg2022_3d,
  title     = {3D Interconnect Architectures for Heterogeneous Technologies: Modeling and Optimization},
  author    = {Bamberg, Lennart and Joseph, Jan Moritz and Garc{\'\i}a-Ortiz, Alberto and Pionteck, Thilo},
  year      = {2022},
  publisher = {Springer International Publishing},
  address   = {Cham},
  doi       = {10.1007/978-3-030-98229-4}
}

\end{document}